\def\be{\begin{equation}}
\def\ee{\end{equation}}
\def\ba{\begin{eqnarray}}
\def\ea{\end{eqnarray}}
\def\l{\label}
\def\n{\nonumber\\}
\def\g{\gamma}
\def\D{\Delta}
\def\a{\alpha_s}
\def\ap{\alpha_P}
\def\bars{\begin{eqnarray*}}
\def\ears{\end{eqnarray*}}
\def\eqar{\end{array}}
\def\beqar{\begin{array}}

\documentclass[12pt]{article}

\usepackage{epsfig}

\begin{document} 
 \title{Unified QCD picture of hard diffraction} 

\author{
H. Navelet and R. Peschanski \\ Service Physique Theorique,
 CE Saclay\thanks{Address: SPhT  CEA-Saclay  F-91191
Gif-sur-Yvette Cedex, France;}  \thanks{emails: navelet@spht.saclay.cea.fr, 
pesch@spht.saclay.cea.fr}}
\maketitle

\begin{abstract}

Using a combination of S-Matrix and perturbative QCD properties in the small 
$x_{Bj}$ regime, 
we propose a formulation of hard diffraction unifying the partonic (Ingelman-Schlein) Pomeron, 
Soft 
Colour Interaction and QCD dipole descriptions. In particular, we show that all 
three approaches give an unique and mutually compatible formula for the proton
diffractive structure functions $F^{Diff}_{T,L}$  incorporating perturbative and 
non perturbative QCD features.

\end{abstract}

\section{ Introduction}

``Hard diffraction'' is an experimental phenomenon which lies at the borderline 
between ``hard'' and ``soft'' interactions. It appears as a scattering process 
initiated by a hard probe (e.g. a virtual photon at HERA \cite{he95} or 
 a forward 
jet at the 
Tevatron \cite{te97}), but in which the proton target is not destroyed, in a 
similar way to 
 conventional soft diffractive processes. It is associated with 
large rapidity gaps between the  hadronic remnants of the target and of the 
projectile as was 
discovered at  
HERA, but was first identified by high-$P_T$ jets in diffraction at the ISR 
\cite{ua8}.

There has been a lot of debate about the theoretical explanation of hard 
diffraction. 
Indeed, assuming a partonic content of the Pomeron led to a nice  prediction 
\cite {in85} of the phenomenon 
and, 
supplemented by the QCD evolution of diffractive structure functions, to a 
quantitative 
description of HERA data \cite {ro01}. As satisfactory as can be this 
phenomenological 
analysis, the main unsolved problem in  this approach is the lack of relation 
between 
diffractive and non diffractive hard scattering leading to a profusion of input 
parameters 
(e.g. the non-perturbative parton distributions in the Pomeron). If one wants 
to enter more 
deeply in the study of hard diffraction and in the still mysterious nature of 
the 
Pomeron 
interaction, one has to look for  theoretical links with QCD. This is 
our goal.

 In the present paper, we shall focus on three existing  different theoretical 
approaches 
of hard diffraction, for which we propose a new, unifying, formulation. The 
first  one we will 
refer to is the ``partonic Pomeron'' approach  \cite 
{in85}.  The 
hard photon  is here supposed to probe the parton distributions of the Pomeron 
Regge pole considered as a hadronic particle.  In fact we 
can also consider it as  
an 
extension of the Regge theory of soft diffraction \cite {ka79} to incorporate 
the effect of  
the hardness of the probe\footnote{Models using explicitely   
concepts of  
the S-Matrix framework of Regge singularity theory and taking into account the 
hard probe on a phenomenological ground do exist \cite{ca95}.}.

 A second approach is the Soft Colour Interaction one , 
where hard diffraction is described by a two-step process: during a relatively 
short 
``interaction 
time'', the probe initiates a typical hard deep-inelastic interaction. Then, at 
large 
times/distances, a soft colour 
interaction 
is assumed which will decide of  the separation between diffractive and 
non-diffractive events according to a simple probabilistic rule: It gives  rise 
to colour 
neutralization of the final state with  probability of order $1/N_c^2$ 
(where $N_c$ is 
the number 
of colours)  and thus to  
rapidity gaps and 
diffraction. The assumed extreme softness of Soft Colour Interaction ensures 
that the dynamics 
of hard 
partons remains unchanged. Various models based on Soft Colour Interaction 
\cite{bu95,in96}  lead 
to 
satisfactory phenomenological descriptions.
 
A third approach is based on the small $x_{Bj}$ regime of perturbative QCD, 
where the 
resummation of leading $\log 1/x_{Bj}$ contributions allows one to obtain some 
theoretical 
information on high energy hard interaction processes following the Balitsky,  
Fadin, 
Kuraev, Lipatov (BFKL) approach \cite{bfkl}. Calculations of hard diffraction  
using the  related QCD 
dipole approach
 \cite{mu94} or  in the original BFKL framework \cite{ba95} have been 
performed.
 In fact, in the line of QCD dipole models for proton structure 
functions \cite{na96}, models \cite {bi96} for   diffractive proton
structure 
functions\footnote{Two different components have to be distinguished 
\cite{ni96,bi96} : {\it 
inelastic} for large diffractive masses and  
{\it elastic} for small masses.} have been derived 
and give a convenient description of HERA data. 
For further  abbreviation we shall call it the ``QCD dipole'' 
approach\footnote{To be 
more specific, we shall not  include either next-leading or non-perturbative 
corrections to 
the 
QCD dipole picture, and use an effective leading 
order BFKL approach.}.

Keeping a general point of view, 
one may notice 
that each of these approaches has advantages and disadvantages with respect to 
the others. The 
partonic Pomeron approach synthetises what we know about the effect of the 
hard probe  and the factorization properties of the hard interaction but gives 
no
prediction about the soft Pomeron dynamics. The Soft Colour Interaction 
approach give a 
nice relation 
(even in normalization) between hard and soft deep-inelastic scattering but the 
nature of the 
colour rearrangement at long distances is unknown. Finally the QCD dipole 
approach gives 
detailed prescriptions for  the form of the amplitudes, but the problem   of the 
perturbative/non-perturbative QCD interface and  the crucial point  
of the 
relative normalization of diffractive {\it vs} non-diffractive contributions, 
remain obscure.

In the present paper we will show that the three aproaches may find a common 
formulation and 
intrinsic mutual equivalence through  S-Matrix relations on different analytic 
discontinuities of  $3 \to 3$ forward elastic amplitudes in the so-called 
triple-Regge regime. Within this unifying S-Matrix framework and  using the hard 
probe as a hint 
for the application of perturbative QCD to the calculation of these 
discontinuities whenever it is reasonably justified, 
we will end 
with a  constrained formulation of the diffractive structure 
functions, transcending 
the limitations of each of them taken separately.

The plan of the paper is as follows: In the next section we will give the 
general framework, 
using both the S-Matrix and perturbative QCD properties, allowing one to 
mutually relate the 
three abovementionned 
approaches to hard diffraction. In the next section 3, concentrating on  
the {\it 
inelastic} component, we show the 
compatibility of the 
QCD dipole and partonic Pomeron approaches. In section 4 the same is done 
for the QCD 
dipole and Soft Colour Interaction 
approaches. As a result, a  determination of the longitudinal and transverse 
diffractive
structure functions emerges incorporating both perturbative and non perturbative 
ingredients in an overall consistent QCD picture of hard diffraction. 
In the last section 5, we give a summary of our results, and an outlook on 
phenomenological tests and theoretical  
consequences of our unifying approach. 

\section{ General unifying framework}

The usual kinematical variables  for the diffractive amplitudes and  
cross-sections are 
$Q^2,$ the 
photon virtuality, $Y\equiv\log 1/x_{Bj},$ the total rapidity available for 
the final 
hadronic state, $y \equiv\log 1/x_{P} ,$ the rapidity gap, $t \simeq -P_T^2,$ 
where $P_T$ is the  
momentum transferred
at the proton vertex. We use also $Y\!-\!y \approx \log 
M_X^2 \simeq \log 1/\beta$ 
where $M_X$ is the 
invariant mass  of the diffractively produced state\footnote{Indeed one has the 
relation $\beta \equiv 
x_{Bj}/x_{P}=Q^2/(M_X^2+Q^2-t)$ but due to the sharp 
cut-off  in t, one can often neglect t in the denominator.}.

In the following we shall make use of the   important S-Matrix  connection, also 
called 
Mueller-Regge relation \cite {mu70}, 
 between 
  semi-inclusive  amplitudes and  specific 
discontinuity contributions of forward elastic $3 \to 3$ amplitudes. It 
naturally applies to hard 
diffraction 
initiated by a virtual photon, as sketched in Fig.1,  namely 
\be
\gamma^* + p\to p + X \ \iff Disc_1\left\{\gamma^* \bar p \ p \ \to 
\gamma^*  \bar p\ p  
\right\} \ 
.
\label{3to3}
\ee
In fact, this relation, derived in the context of soft physics\footnote{To our 
knowledge, the 
application of the Mueller-Regge relations for the Soft Pomeron region  appeared 
in a Regge theory framework  
in Refs.\cite{dt70}, while the phase factors were discussed in detail in 
\cite{dt71}. The extension to soft diffraction and related references can be 
found in 
Ref. 
\cite{ka79}.} 
\cite{dt70}, can also be applied to the 
definition 
of hard 
diffraction when $Q^2$  and $y \equiv\log 1/x_{P}$ are large enough 
(typically, $Q^2> 4.5 GeV^2, x_{P}> 
10^{-2}).$ In this case, the Mueller-Regge relation applies in a region where 
the Pomeron Regge 
poles are 
supposed to take into 
account the colour singlet exchange responsible for the gap. If moreover, the 
diffractive mass $M_X$ 
is large enough, one considers  a 
Pomeron-like singularity describing the large $M_X$ behaviour. However the 
resulting 
triple-Pomeron contribution is to be interpreted here in a loose sense since  
this third Regge
singularity may differ from the other ones due to the hard probe, as mentionned 
in the previous section. We will thus keep the word  Pomeron for 
those responsible for the gaps.

In the triple-Regge kinematical region, one thus writes \cite{ka79}: 

\be
\frac {dF_{T,L}^{Diff}}{dP_T^2} \sim Disc_1 A(3 \to 3) \ \propto \ G_P(t) \ 
\vert\xi_{\ap}\vert^2 \ 
e^{2(\ap -1) y} \ \sigma^{tot}_{\gamma^*-P}\ ,
\label{Regge}
\ee
where $\frac {dF_{T,L}^{Diff}}{dP_T^2}$ is the ``unintegrated'' (t-dependent)
diffractive structure 
functions (related to the measurable hard diffractive cross-sections), 
\be 
\xi_{\ap}(t) \equiv \Gamma(-\ap (t)) \left\{1+e^{-i\pi\ap (t)}\right\}
\label{phase}
\ee
is the (partonic) Pomeron Regge phase factor in the amplitude and 
${\ap}(t) = \ap(0) + {\ap^{\prime}} t $
is the  (partonic) Pomeron-pole Regge trajectory\footnote{Other Regge 
singularities, such 
as cuts, 
could appear and give logarithmic prefactors. For sake of simplicity, but 
without affecting the 
conclusions, we 
will stick to the formalism of ``effective'' Regge 
poles.} with intercept $\ap(0)$ and slope 
$\ap^{\prime}$.  $G_P(t)$ is a function, not constrained in the Regge formalism, 
describing the coupling of incident Pomerons to the 
proton.

The Pomeron-photon cross-section $\sigma^{tot}_{\gamma^*-P}$ describes the 
interaction of the hard probe with the incident Pomeron. For instance, in the 
partonic Pomeron model \cite {in85} , it is expressed in terms of Pomeron 
structure functions, in much the same way as proton-photon cross-sections
are expressed in terms of the proton structure functions, but with a quite 
different partonic 
content \cite{ro01}.

\begin{figure}[t]
    \centerline{\epsfig{figure=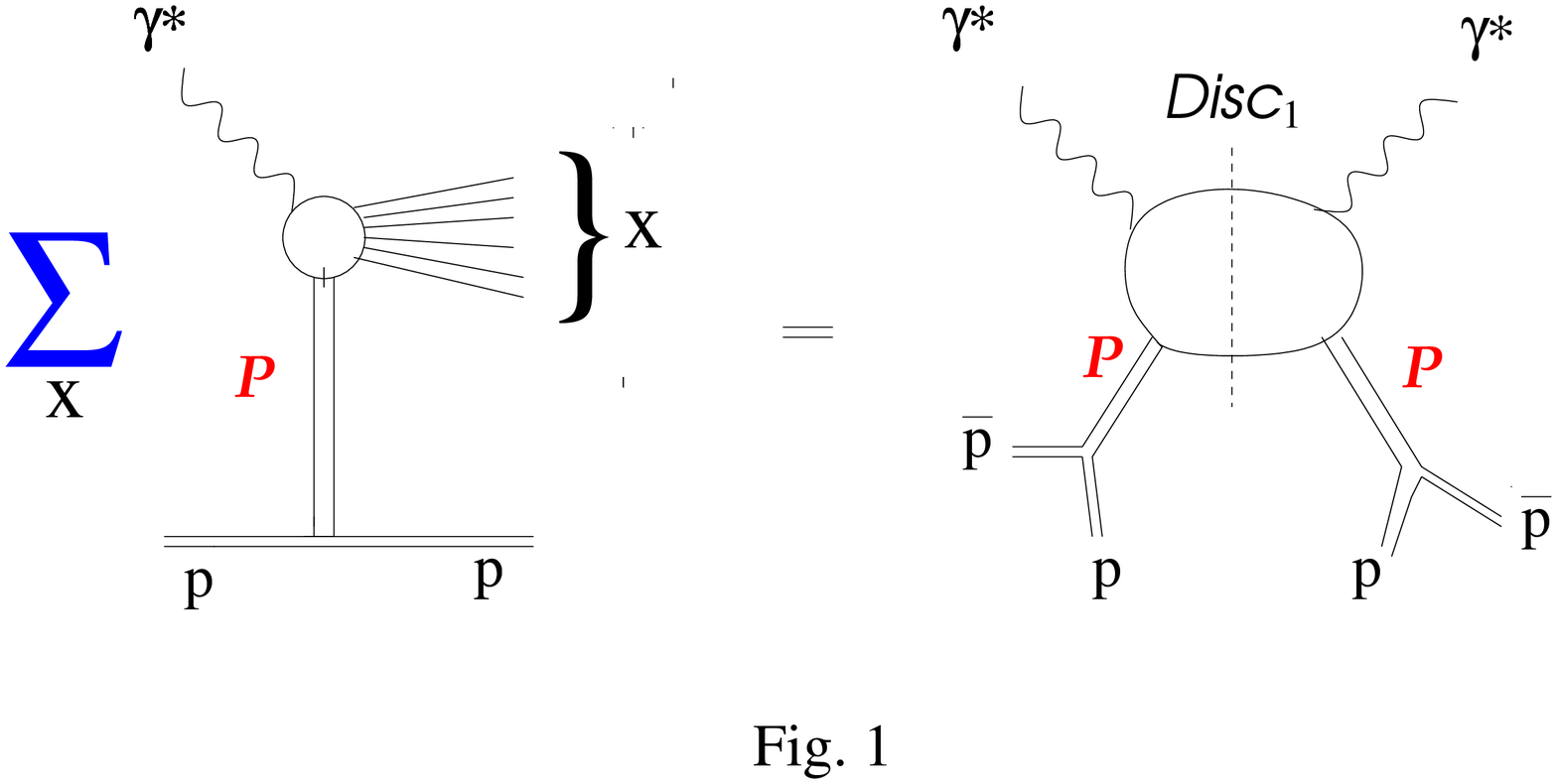,width=10cm}}
    \caption{ {\footnotesize
{\bf Mueller-Regge relation for hard diffraction.} 
The summation 
over final diffractive states {\bf X} gives rise to a specific discontinuity 
$Disc_1 A$ of a 
forward  $A(3 \to 3)$ amplitude. {\cal P} denote the Pomeron-Regge exchange 
leading to 
rapidity gaps
     .}}
\end{figure}

Quite interestingly, the existence of the Regge phase factors (\ref{phase}) 
allows one  to  relate 
other discontinuities of $A(3 \to 3)$ to $Disc_1 A.$   As sketched in 
Fig.2, one may 
consider a double discontinuity $Disc_2 A(3 \to 3)$ taking into account also the 
analytic discontinuity  in the subenergy of one of the incident
 Pomeron exchanges 
and then a triple discontinuity $Disc_3 A(3 \to 3)$ including the discontinuity 
over the two 
Pomeron 
exchanges. The 
expression of the 
discontinuities, through generalized unitarity relations, is obtainedthrough the 
imaginary part of the 
relevant Regge phase 
factors. One writes\footnote{The derivation can be found in the Physics Reports 
of Ref. \cite{dt71} with the complete expressions in page 319 for the cut and 
uncut  triple-Regge vertices. The main issue is that the single discontinuity 
$Disc_1$ eliminates already all terms but one in the rather involved expression 
of the uncut vertex.} \cite{dt71}:
\be
Disc_1 A \! = \!Disc_2 A\! = \!Disc_3 A =
\frac{\sin \pi (2\ap(t)\!-\!\ap^*(0))}{\sin \pi \ap^*(0)}\ {\cal V}_{p\bar p}\ ,
\l{disc}
\ee
where ${\cal V}_{p\bar p}$ ia a real function describing the  
$p\bar p$ vertex in 
$A(3\! \to \!3).$ The notation $\ap^*(0)$ indicates that the third Regge
exchange, corresponding to the summation over diffractive final states, 
may have (and indeed {\it has} in our calculations) a different effective 
trajectory\footnote{ In particular,  $\ap^*(0) \ne 1,$ avoiding a definition 
problem in (\ref{disc}).}.

\begin{figure}[t]
    \centerline{\epsfig{figure=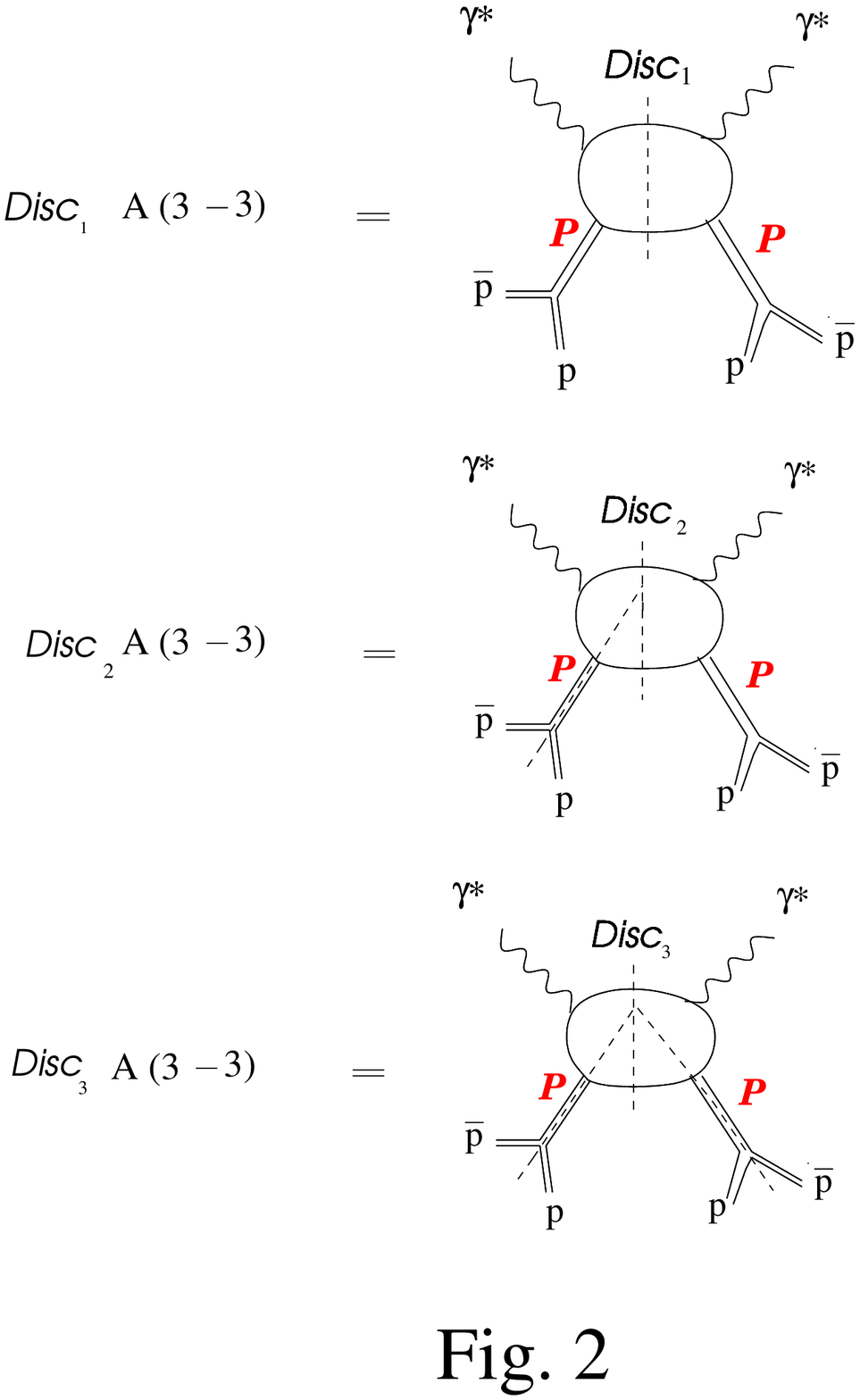,width=6cm}}
    \caption{ 
{\footnotesize {\bf Three analytic discontinuities of 
$A(3 \to 3).$ } The single $Disc_1 A,$ double 
$Disc_2 A,$ and triple  $Disc_3 A$  discontinuities of the 
triple-Regge $A(3\! \to \!3)$ amplitude are displayed
     .}
}
\end{figure}

Beyond these general relations of the S-Matrix framework in the triple-Regge 
regime, 
we will now take 
advantage of the hard probe, allowing one to introduce in the game the 
(resummed) 
perturbative QCD expansion at 
high energy (small $x_{Bj}$). 

Let us start with general considerations. In a generic S-Matrix approach, the 
analytic discontinuities of scattering amplitudes are related to a 
summation over a complete set of asymptotic {\it hadronic} 
final states. If however, 
the underlying microscopic field theory is at work with small renormalized 
coupling constant due to 
the hard probe, it is possible in some cases to approximate the same 
discontinuity using a 
complete set of {\it 
partonic} states. In particular, at high energy and within the approximation of 
leading logs (and also 
large $N_c$), QCD dipoles can be identified  as providing such a basis \cite 
{mu94}. The interest 
is that  the wave 
function of incoming states at 
the hard vertex is completely determined at the leading log level. Using this 
wave-function\footnote{In fact only $1\to 1$ and $1\to 2$ dipole transitions are 
needed \cite{mu94,bi96}.}
together with 
the 
appropriate $k_T$ 
factorization properties at the proton vertices \cite{ca91}, one is able to 
estimate the  contributions to 
the various 
discontinuities of $A(3\! \to \!3)$ depicted in Fig.2 within the (resummed) 
perturbative QCD framework.

Hence, the S-Matrix relations (\ref{disc}), which express a 
simple phase relation in a pure Regge framework, acquire a non-trivial meaning 
if 
one considers the various perturbative QCD 
 ingredients and non-perturbative interfaces for the 
different discontinuities. In 
some sense, at it 
is shown in Fig.3, the discontinuities can be associated, whenever it is 
possible, with the hard QCD interaction at 
some time $T \approx 1/Q,$ supplemented by a perturbative resummation of QCD 
radiation  around this interaction time. The perturbative/non-perturbative 
interface has to be taken into account in order to put a reasonable limit to 
this derivation. Depending on the discontinuity considered, 
one has different 
interaction pictures with a hard interaction at short time, and soft  
interaction 
diluted in space-time. Instead of considering these pictures  as different, we 
will use the S-Matrix relations as  a link between  these 
pictures 
and thus will get non-trivial relations, as we will see later on. Let us comment  
in 
turn on the pictures 
corresponding to $Disc_1, 
Disc_2 , Disc_3.$

\begin{figure}[t]
    \centerline{\epsfig{figure=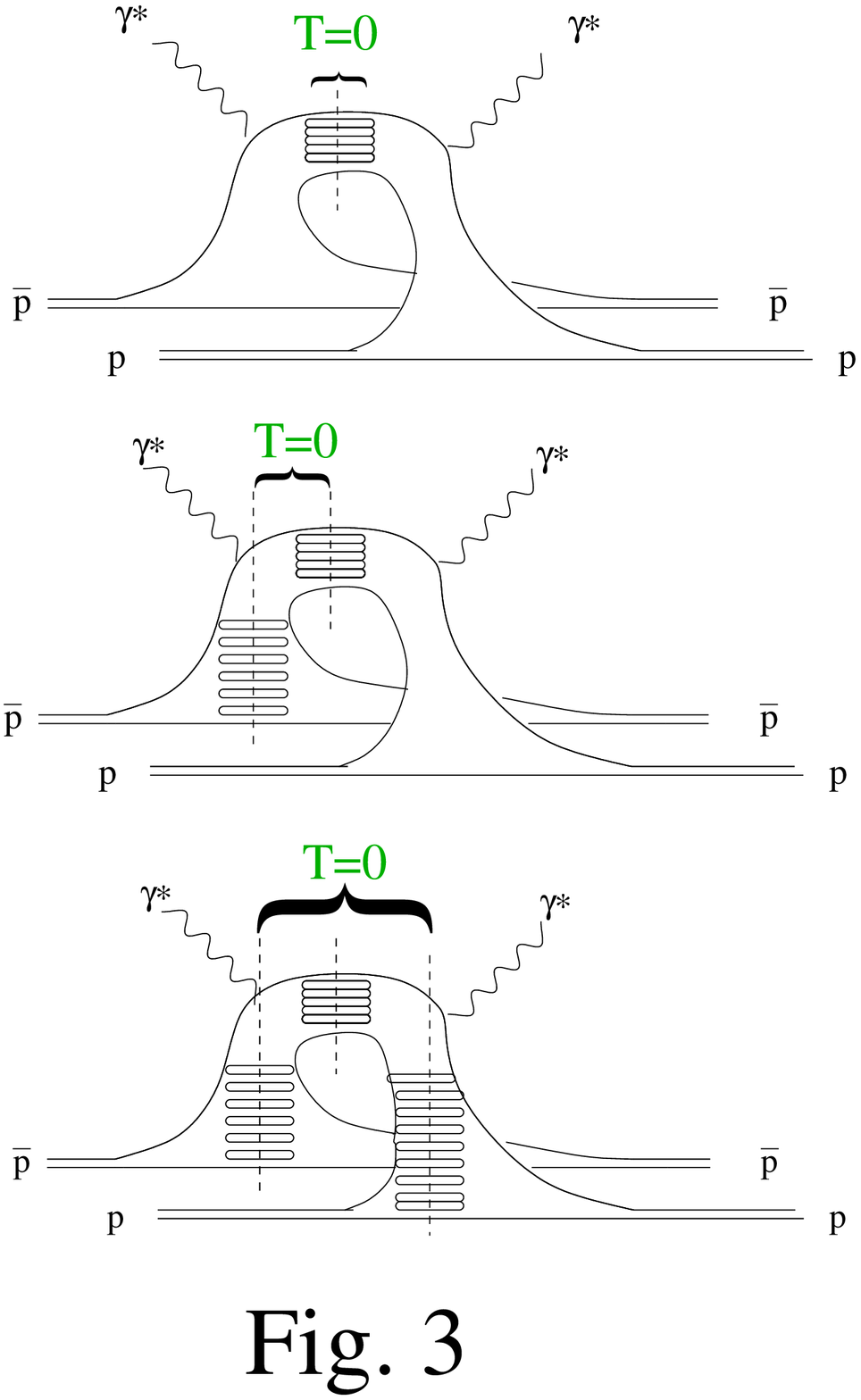,width=6cm}}
    \caption{ {\footnotesize
{\bf ``Time-dependent'' picture of the 
$A(3 \to 3)$ discontinuities.}
Upper graph: Description of $Disc_1 A(3 \to 3)$ (partonic Pomeron 
approach); Middle 
graph: Description of $Disc_2 A(3 \to 3)$ (candidate for the Soft Colour 
Interaction approach); Lower graph: 
Description of $Disc_3 A(3 \to 3)$ (QCD dipole approach). 
     }}
\end{figure}

{\bf i)} $Disc_1,$ as we have seen from relation (\ref{Regge}), is parametrized 
in 
terms of soft Pomeron trajectories and of $\sigma^{tot}_{\gamma^*-P},$  the 
unknown quantity  
parametrizing the interaction of the incident Pomeron with   the hard probe. The 
same  $\sigma^{tot}_{\gamma^*-P}$ is the one 
described by the 
partonic structure functions of the Pomeron in the original scheme of 
Ref.\cite{in85}.  $Disc_1$  is thus a natural  framework for the partonic 
Pomeron approach, since the analytical discontinuity over the diffractive final 
states is effectively expressed in terms of   the hard interaction of the photon 
with the partons inside the incident Pomeron. This is schematically depicted in 
the ``time-dependent'' scheme of Fig.3, where the soft incident Pomerons are 
represented 
as living during a ``long'' time, while the partonic (or QCD dipole) process is 
initiated by the hard interaction.

{\bf ii)} $Disc_2$ is quite interesting since it appears as a good candidate for
a description of the Soft Colour Interaction approach, see Fig.3. It appears as 
a partonic 
interaction very similar to 
the one describing ordinary deep-inelastic processes, in parallel with a 
``soft''  
correction evolving during a long time, corresponding to the uncut Pomeron 
singularity in Fig.3. In  section 
4, we will 
see how this can be 
quantified using the interrelation with the QCD dipole picture.

{\bf iii)} $Disc_3:$ In this case the discontinuity is the ``full'' one, 
cutting the two Pomerons 
together with the $\gamma^*\!-\!P$ interaction. Indeed, this full cutting is 
required by 
consistency of the QCD 
dipole picture based on a QCD calculation 
of the partonic 
states at the interaction time \cite{mu94}. In the description of hard 
diffraction (see, e.g.,  
\cite {mu94}, the amplitude is deduced 
from a $1 \to 2$ 
dipole transition, starting either from the primordial $q \bar q$ state of the 
photon (``elastic 
component'') or from a dipole excited in the photon wave function for higher 
diffractive masses
(``inelastic component''). In the following we shall focus on the main 
``inelastic component'',
leaving the ``elastic component'' for a further study\footnote{We expect the 
``elastic component'' to be related to higher twist terms in an operator 
product 
expansion of diffractive structure functions
\cite{ba98}.}.

This last analysis requires some more care about the 
perturbative/non-perturbative interface, which cannot be restricted to usual 
factorisation properties. As studied  in Ref. \cite{ba96}, one has to 
distinghuish the cases whether we consider strictly the forward 
$t=0$ case or not. Indeed, whenever $t \ne 0,$ the QCD radiation accompanying 
the hard probe extends far into the infrared region  at the triple Pomeron 
vertex, making the perturbative treatment difficult. At $t=0,$ due to an 
interesting conservation law of conformal dimensions in the BFKL framework, the 
perturbative resummation acquires more credibility, at least in the same 
approximation as for the total structure functions,. We will thus stick to using 
the perturbative 
calculations at $t=0$,  more precisely inspiring ourselves from those performed 
in paper Ref. \cite{bi98} in the 
framework of the QCD dipole model\footnote{These calculations are similar to 
those performed in the BFKL framework \cite{ba96}, with the simplification that 
the 4-gluon intermediate state does not appear 
in the $1/N_c$ limit of the QCD dipole model. The inclusion of such a state will 
be welcome, especially for large diffractive masses, but lies beyond the scope 
and the dipole framework of our present paper.}.

Due to the relations (\ref{disc}), we expect that all three 
descriptions of hard 
diffraction, even if having seemingly distinct ``time evolutions'', indeed 
correspond to 
an unique one in a complete QCD theory 
satisfying both macroscopic 
(S-Matrix) and microscopic (with interacting quarks and gluons) requirements. In 
the 
absence yet of such 
a complete realization of the strong interaction theory, we shall adopt the 
following point of 
view: looking for a  synthesis, we will first check whether all three  
approaches are consistent
between themselves and then formulate hard diffractive structure functions 
realizing the synthesis.

\section{From  QCD dipoles to partonic Pomeron}

Our  starting point   is the formula for the differential structure functions 
(inelastic component) 
 at $P_T \!=\!0$ for longitudinal and 
transverse photon given in \cite{bi98}:
\ba
\frac {dF_{T,L}^{Diff}}{dP_T^2}(Q^2,Y,y;P_T\!=\!0) = \ \ \ \ \ \ \ \ \ \ \ \ \ \ 
\ \ 
\ \ \ \ \ \ \ \ \ \ \ \ \ \ \ \ \ \ \ \ \ \ 
 \n =\frac {{\cal N}^{Diff}}{x_P Q_0^2}\ 
\int_{c-i\infty}^{c+i\infty}\frac 
{d\g_1}{2i\pi}\frac {d\g_2}{2i\pi}\frac {d\g}{2i\pi}\ \delta 
(1\!-\!\g_1\!-\!\g_2\!-\!\g)\ 
\Theta P\!f_{T,L} \n
\times \left(\frac Q{Q_0}\right)^{2\g}\ \exp \left\{y [\D(\g_1)+\D(\g_2)] + 
(Y\!-\!y)\ 
[\D(\g)]\right\}\ ,
\l{biczy}
\ea
where
\be
\D(\g) \equiv \frac {\a N_c}{\pi} \chi (\g) = \frac {\a 
N_c}{\pi}\left\{2\psi(1)-2\psi(\g)-2\psi(1-\g)\right\}
\l{chi}
\ee
 is the BFKL evolution kernel \cite{bfkl}, $Q_0$ the non-perturbative scale of 
primordial 
dipoles in the proton \cite{na96},
\be
{\cal N}^{Diff}\equiv \frac {2\a ^5\ N^2_c e_f^2}{\pi^3}\   
n^2_{eff},
\l{nordif}
\ee
the normalization with $n_{eff}$ the phenomenological number of primordial 
dipoles in the proton at scale 
$Q_0$ and $e_f^2$ the sum over squared quark charges. The other prefactors of 
perturbative QCD origin are:
\be
P\!f_{T,L}\equiv hG(\g_1)\ hG(\g_2)\ I_{T,L}(\g)\ G^{-1}(\g)\ ,
\l{Pf}
\ee
with
\be
h(\g)=\frac 1 {4\g^2(1-\g)^2}\ ;\ G(\g)= \frac {\Gamma(\g)}{\Gamma(1-\g)}\ ,
\l{hG}
\ee
and finally,
\bars
\left( {\beqar{c} I_L \\ 
 I_T \eqar} \right) \equiv  
\frac  1{\gamma\ (1-\gamma)}\ 
\frac{\Gamma^2(1+\g)\Gamma^4(2-\g)}{\Gamma(4-2\g)\Gamma(2+2\g)} 
\ \left( 
{\beqar{c} {2\gamma\ (1-\gamma)} \\
{(1+\gamma)(2-\gamma)}\eqar}\right)\ ,
\label{ITL}
\ears
which are  the ` `impact factors'' \cite{bfkl,ca91,na96} of the longitudinal and 
transverse photon in 
the BFKL formalism.
$\Theta$ is  a 
cut-off 
function taking into account the non-perturbative description of the end of the 
rapidity 
spectrum 
of emitted gluons \cite {bi98}. In practice, it mainly affects the overall
normalization (also the form at edges of the $\beta$-dependence). It is one 
aspect of 
the lack of predictivity of the QCD dipole approach about the relative 
normalization of 
diffractive vs. non diffractive contributions.

One important ingredient of formula (\ref{biczy}) is the delta function $\delta 
(1\!-\!\g_1\!-\!\g_2\!-\!\g).$ It expresses the conservation law of conformal 
dimensions at the triple pomeron vertex, which is a special feature of the BFKL 
property in the forward direction. As discussed in the previous section, we have 
choosen the QCD dipole formulation for the forward diffraction 
instead of the non-forward one \cite{mu94}, since  it is expected to be less 
affected by non -perturbative contributions. On the other hand, it is 
phenomenologically not 
difficult to insert the non-perturbative phenomenological information on the 
$t$-dependence coming from the proton vertices in the non-forwardregion. Indeed, 
it is convenient to parametrize the integrated 
diffractive structure function assuming an exponential behaviour $\frac 
{dF_{T,L}^{Diff}}{dP_T^2} \approx \frac {dF_{T,L}^{Diff}}{dP_T^2}(P_T\!=\!0)\ e^
{-P_T^2/<P_T^2>},$ well verified experimentally\footnote{The mean squared 
transverse 
momentum  $<P_T^2>$ is presently known 
\cite{ZE97} only on a global average to be of order $7 GeV^2.$ It could depend 
on  other 
variables but  it happens to be 
very similar to the diffractive photoproduction slope, confirming the stability 
of this non perturbative parameter in different kinematical regions. We shall 
assume it 
constant in 
the following.
} \cite{ZE97}. We will thus in the following assume the relations:
\be
F_{T,L}^{Diff}(Q^2,Y,y) \equiv \int dP_T^2\ \frac {dF_{T,L}^{Diff}}{dP_T^2} 
\approx 
\ <P_T^2>\frac {dF_{T,L}^{Diff}}{dP_T^2}(P_T=0).
\l{Fdiff}
\ee
 
The first step of the computation of formula (\ref{biczy}) is \cite{ba95,bi98} 
to use the 
saddle-point 
approximation 
at large $y$ to integrate over the difference $\g_1-\g_2.$ One easily gets
\ba
\frac {dF_{T,L}^{Diff}}{dP_T^2}(P_T\!=\!0)= 
{\cal N}^{Diff}\frac 1{x_P Q_0^2}\ \int_{c-i\infty}^{c+i\infty}\frac 
{d\g}{2i\pi}\sqrt {\frac 
1 {4\pi \D ^{''}\! (\frac {1-\g} 2)\ 
y}} 
\n  \Theta P\!f_{T,L} \ 
\exp \left\{2y [\D\left(\frac {1\!-\!\g} 2\right)] + (Y\!-\!y)\ [\D(\g)] + 2\g 
\log\frac 
Q{Q_0}\right\} \ .
\l{biczy1}
\ea

For comparison the QCD dipole 
formula for the total proton structure functions \cite{na96} is:
\be
F_{T,L} = {\cal N}^{tot}\ \int_{c-i\infty}^{c+i\infty}
\frac {d\g}{2i\pi}
\ e^{Y\D(\g)}\ \left(\frac Q{Q_0}\right)^{2\g}\ h(\g)\ I_{T,L}(\g),
\l{tot}
\ee
with 
\be
{\cal N}^{tot} = \frac {\a ^2 N_c}{\pi} \ n_{eff} \ e_f^2\ .
\l{normtot}
\ee

In order to proceed further in our theoretical analysis, it is suitable to look 
for an  
analytical solution of (\ref{biczy1}). For 
this sake we shall  now use another saddle-point approximation  near $\g =0.$ In 
previous analyses, the integral was either approximated in some regions of the 
variables \cite{ba95} or numerically evaluated \cite{bi98}. For sake of a 
theoretical discussion in parallel with the formula  (\ref{tot}) for the total 
structure functions, we better choose  a gaussian approximation valid in a 
reasonably large physical region.

The gaussian 
approximation makes use of the following expansions
\ba
\D\left(\frac {1\!-\!\g} 2\right) &\approx& \D + \frac {\D ^{''}}8\g^2  + {\cal 
O}(\g^4)\n
\D(\g) &\approx& \D + \frac {\D ^{''}}2\  \left(\frac 12-\!\g\right)^2 + {\cal 
O}\left((\frac 12-\!\g)^4\right)\ ,
\l{expan}
\ea
where by definition 
\be
\D \equiv \D(\g\!=\!\frac 12) = 4\log 2\ \frac {\a N_c}{\pi}  ; \D ^{''}\equiv 
\D^{''}(\g\!=\!\frac 12) 
 =28 \zeta (3)\ \frac {\a N_c}{\pi} \ .
\l{zeta}
\ee
Note that  $0<\gamma < 1/2,$ which allows one to use the gaussian 
saddle-point for  diffraction in a  region  avoiding the edges of 
integration\footnote{We checked the approximate validity of the gaussian 
approximation by comparison with a non-Gaussian one $\D(\g) \approx \frac {\a 
N_c}{\pi} 
\left(\frac 1{\g} +  4\zeta(3)\  \gamma^2\right)$ valid when $\g \sim 0$ 
\cite{ba95} and 
with numerical 
estimates 
\cite{bi98}. More precisely,  $4\log 2 \approx 2.77;\ 28 \zeta (3) \approx 33.6\ 
,$  giving $\D ^{''}/8\D$
of order 1. Hence the approximations  (\ref{expan}) will be good for 
$\D\left(\frac {1\!-\!\g} 2\right),$  less for $\D(\g),$ but enough for our 
theoretical purpose.} $\gamma =0, \frac 12.$ .

It is important to realize here a different expected result with the gaussian 
approximation for the total structure function (\ref{tot}) in the BFKL formalism 
where one 
uses a 
saddle-point approximation near $\g = 1/2.$ In particular, 
the 
dynamical exponent $\D(\g)$ will be moved away from its canonical value  at 
$\g = 1/2,$ even without  the  virtuality factor $\left(\frac 
Q{Q_0}\right)^{2{\bf \g}}.$

After using the gaussian approximation, and relation (\ref{Fdiff}), the 
resulting 
structure functions 
read:
\ba
F_{T,L}^{Diff}(Q^2,Y,y) = 
{\cal N}^{Diff}\frac {<P_T^2>}{x_P Q_0^2}\ \Theta\ P\!f_{T,L}\frac {\exp 
(2y\D)}{4\pi \D 
^{''} y}
\n \times\sqrt {\frac 2 
{1+2\eta}}
\exp \left\{ (Y\!-\!y)\ {\bf \epsilon}   _s \right\}\ \left(\frac 
Q{Q_0}\right)^{2{\bf \g_s}}
\exp \left( - \frac {2 \eta}{1+2\eta}\ \frac {2\log^2\left(\frac 
Q{Q_0}\right)}{\D ^{''} 
(Y\!-\!y)}\right) \ ,
\l{biczy2}
\ea
where the saddle-point determines both  the new ``effective dimension'' and 
``effective 
intercept'':
\be
{\bf \g_s}= \frac {\eta}{1+2\eta}\ ;\ 
{\bf \epsilon}_s  =\D + \frac {\D^{''}}{8 (1+2\eta)}
\l{epsilon}
\ee
for the Pomeron-photon cross-section.
By definition
\be
\eta = \frac {Y\!-\!y}{y} = \ \frac {\log1/{\beta}}{\log 1/ {x_P}}\ . 
\l{eta}
\ee

The connection between the QCD dipole model and the partonic Pomeron is thus 
well 
illustrated by 
the identification of our resulting formula (\ref{biczy2}) with the triple Regge 
prediction 
(\ref{Regge}) considered at $t=0.$ Indeed,  We identify the partonic 
Pomeron 
intercept as
$\ap(0) \equiv 1+\D,$ which is natural in a BFKL framework at $Q=Q_0,$ see the 
following discussion. We get now the 
following new relation:
\be
\sigma^{tot}_{\gamma^*-P} \sim \sqrt {\frac {2}
{1\!+\!2\eta}}
\  {\exp \left\{ (Y\!-\!y) {\bf \epsilon}   _s \right\} }\left(\frac 
Q{Q_0}\right)^{2{\bf \g_s}}
\!\exp \left( -\frac  {2\eta}{1\!+\!2\eta} \frac {2\log^2\left(\frac 
Q{Q_0}\right)}{\D 
^{''} (Y\!\!-\!y)}\right) \ ,
\l{biczy3} 
\l{QCDRegge}
\ee
with (\ref{epsilon}) defining the ``effective'' anomalous dimension  and 
intercept of a  ``deformed'' BFKL formula for the photon Pomeron (with zero mass 
$=\sqrt 
{-t}$) total cross-section.

It is interesting to compare expression (\ref{biczy2}) and ${\bf \g_s},{\bf 
\epsilon}  _s$ with 
the 
canonical BFKL formula (\ref{tot}) in the saddle-point approximation namely:
\be
F_{L,T} \approx  {\cal N}^{tot}\ h\ I_{T,L}\ 
\frac {\exp ({Y\D})}{\sqrt {2\pi\D ^{''} \ Y}}\ \left(\frac Q{Q_0}\right) \ \exp 
\left( - 
\frac 
{2\log^2\left(\frac Q{Q_0}\right)}{\D ^{''} \ Y}
\right) \ .
\l{tot2}
\ee

The ``effective BFKL'' parameters for 
$\sigma^{tot}_{\gamma^*-P}$
differ from the BFKL  ones by an amount depending on $\eta = 
\frac {Y\!-\!y}y.$ Also note that the ``diffusion'' term, which quantifies the 
amount of 
transverse momentum drift of gluons (or dipoles) along the BFKL evolution \cite 
{ba93} is 
renormalized by 
a 
factor $2\eta/(1+2\eta),$ the same as for the anomalous dimension ${\bf \g_s}.$ 

While at large $\eta$ the formulae become similar, since for instance ${\bf 
\g_s} \to \frac 
12, {\bf \epsilon}_s \to \D,$ they sizeably differ from them provided 
$\eta 
\ge {\cal O}(1).$ In fact, this is realized in practice, since in present 
experiments 
$\beta \ge 
x_P.$ On contrary, the features of the Pomeron exchanges responsible for the 
gaps are those of  the ``bare'' BFKL values taken at $Q\equiv Q_0,$ which can be 
interpreted as the partonic Pomerons in the BFKL formalism since the scale $Q_0$ 
is characteristic of 
the 
starting point of the QCD evolutions.

 Thus the ``deformed'' BFKL parameters (\ref{epsilon}) 
lead to a 
situation which can be called intermediate between a ``hard'' (with effective 
dimension $\g= \frac 12$) and a ``soft'' (with $\g= 0$) Pomeron, depending on 
the 
ratio $2\eta/(1+2\eta).$ Also, the energy dependence is faster for the 
``deformed'' BFKL singularity than for the partonic Pomeron. This original 
situation is confirmed by numerical 
estimates \cite{bi98}. 

\section{From dipoles to Soft Colour Interaction}

In order to confront and unify the QCD dipole and Soft Colour Interaction 
approaches, we will  estimate the 
overall contribution of hard diffraction to the total structure function at 
fixed value 
of $x_{Bj}.$ As previously done, we will stick here to the ``inelastic'' 
(leading-twist) 
diffractive component leaving for further study the ``elastic'' component with 
small 
diffractive masses. 

Let us then consider the following integral:
\ba
F_{T,L}^{Diff/tot}\!\! &=& \!\int_{x_{Bj}}^{x_{gap}} dx_PdP_T^2 \ \frac  
{dF_{T,L}^{Diff}}
{dP_T^2}(Q^2,Y,y;P_T) \n &\approx& 
\!\int_{x_{Bj}}^{x_{gap}}  dx_PdP_T^2 \ <P_T^2>
\frac {dF_{T,L}^{Diff}}{dP_T^2}(Q^2,Y,y;P_T=0)
\n
&=& \! \frac {{\cal N}^{Diff}<P_T^2>}{Q_0^2}
\int_{c-i\infty}^{c+i\infty}\frac 
{d\g_1}{2i\pi}\frac {d\g_2}{2i\pi} \frac {d\g}{2i\pi}\delta 
(1\!-\!\g_1\!-\!\g_2\!-\!\g)\ 
\Theta P\!f_{T,L} \n
&\times& \!\left(\frac Q{Q_0}\right)^{2\g}\!\int_{y_{gap}}^Ydy\ e^{\left[y 
\left(\D(\g_1)+\D(\g_2)\right) + (Y\!-\!y)\ 
\D(\g)\right]}\ ,
\l{biczy4}
\ea
where $y_{gap}\equiv \log\frac 1{x_{gap}}$ is the 
minimal value retained for 
the rapidity 
gap. Indeed, the integral over $y$ is expected to be dominated by the behaviour 
of the 
integrand for large enough gaps. We also keep the 
approximation\footnote{See 
the discussion at the end of section 2 and footnote 12.} 
$<P_T^2>=cst.$ 

Technically, the integrals in (\ref{biczy4}) can be performed using a 
saddle-point method in both variables $\g$ and $y$ provided the conditions
\be
\D \ Y \gg 1\ ;\ \frac {\log\frac Q{Q_0}}{ \D^{''} \ Y}\le 1
\l{conditions}
\ee
are realized. The saddle-point equations, see Appendix {\bf A2}, lead to 
remarkably simple solutions
\ba
y_c &=& \left(Y+\frac {2\log\frac Q{Q_0}}{\Delta^{'}(\g_c)}\right)
\ \left(1+\frac {\Delta^{'}\left(\frac{1-\g_c}{2}\right)}
{\Delta^{'}(\g_c)}\right)^{-1} 
\n
\D(\g_c)&=&2\D\left(\frac{1-\g_c}{2}\right)\ \,
\l{solution}
\ea
resulting in a value of $\g_c\simeq 0.175 $ which is ``universal'', i.e. 
independent of the 
kinematics of the reaction.

 The  saddle-point calculation  leads to the following expression:
 \be
F_{T,L}^{Diff/tot}\! =\!   
{\cal N}^{Diff}\frac {<P_T^2>}{Q_0^2} 
\frac {\Theta P\!f_{T,L} }
{\vert\D^{'}(\frac {1\!-\!\g_c}2)\!-\!\D^{'}(\g_c)\vert}
\left(\frac Q{Q_0}\right)^{2\g_c}  \frac{\exp \left(Y 
\D(\g_c)\right)}{\sqrt{4\pi\D^{''}\left(\frac{1-\g_c}{2}\right)y_c}}
\ .
\l{biczy7}
\ee
Note that   $\D^{'}(\frac {1\!-\!\g_c}2)\!+\!\D^{'}(\g_c) <0$ when $0<\g <\frac 
12.$

Let us now come to the main outcome of this calculation. Let us compare the 
result
(\ref{biczy7})  to a {\it total} hard contribution 
of BFKL 
type, 
see equation (\ref{tot}), which can be rewritten in a saddle-point 
approximation:
\be
F_{L,T} \approx  {\cal N}^{tot}\ h(\g_{s.p.})\ I_{T,L}(\g_{s.p.})\  \left(\frac 
Q{Q_0}\right)^{2\g_{s.p.}}
\frac {\exp \left(Y\D (\g_{s.p.})\right)}{\sqrt {2\pi\D ^{''} \ Y}}  \ ,
\l{tot1}
\ee
with the BFKL saddle-point value 
\be
\g_{s.p.} = \frac 12 \left( 1-  4\frac 
{\log\left(\frac Q{Q_0}\right)}{\D ^{''} \ Y}
\right)\ .
\l{value}
\ee
  Hence, (\ref{biczy7}) 
is strikingly similar to  a  BFKL formula (\ref{tot}) with a shift in the   
value of the 
saddle-point namely:
\be
\g_{s.p.}\Rightarrow\g_c \ . 
\l{shift}
\ee

It is  remarkable that the total diffractive contribution to the structure 
function has 
the same analytical form as the non-diffractive one, up to the  substitution 
(\ref{shift}) and the prefactors, which we will study next. This QCD derivation 
is thus compatible with the  postulate of 
the 
Soft Colour Interaction approach, 
namely the  relation between the diffractive and  non diffractive deep-inelastic 
processes stating that the diffractive part of the total hard cross-section is 
not 
intrinsically different. In the SCI framework, this is explicitely realized as a 
whole 
in the first paper of Ref. \cite{bu95} while it is realized  graph by graph in 
the 
approach of paper
\cite{in96}.  In the framework of our QCD calculations, it is however realized 
in a 
somewhat modified way, through a shift 
(\ref{shift}) in the ``effective'' parameters of the hard interaction.
We 
will comment later upon possible phenomenological consequences of this effect.
 
Using our identification of the  hard interaction of BFKL type present in 
$F_{T,L}^{Diff/tot},$ we are now able 
to take advantage of the second  postulate of the Soft 
Colour Interaction approach, i.e. the probabilistic evaluation of the soft 
rearrangement 
of colour at large distances, leading to the famous factor
 $1/N_c^2$ in the ratio of   diffractive vs.  non 
diffractive cross-sections. This 
beautifully 
simple 
proposal  expected from non perturbative QCD properties 
can be 
incorporated in the unified approach in an easy way. We 
are thus led to  a proportionnality between the prefactors of the BFKL 
expressions 
(\ref{tot2}) and
 (\ref{biczy7}), namely
 \be
   \frac {{\cal N}^{Diff}<P_T^2>}{Q_0^2}\ 
\frac {\Theta P\!f_{T,L} (\g) }
{\vert\D^{'}(\frac {1-\g}2)+\D^{'}(\g)\vert}\ \iff  {\bf \frac 1{N_c^2}}\ 
{\cal 
N}^{tot}\ 
h(\g)\ I_{T,L}(\g)\ .
 \l{ident}
 \ee

Using our result (\ref{biczy7}) and assuming 
thus a relation between the prefactors compatible with (\ref{ident}), the 
total diffractive contribution can be rewritten
 \be
F_{T,L}^{Diff/tot}  \approx    {\bf \frac 1{N_c^2}}\ {\cal 
N}^{tot}\  h(\g_c) \ I_{T,L}(\g_c)\ 
\left(\frac Q{Q_0}\right)^{2\g_c}  \frac{\exp \left(Y 
\D(\g_c)\right)}{\sqrt{4\pi\D^{''}\left(\frac{1-\g_c}{2}\right)y_c}}\ 
.\l{biczy9}
\ee

Coming back now to our saddle-point result (\ref{biczy2})
for the differential hard diffractive structure 
function 
(\ref{biczy}), one rewrites
\ba
F_{T,L}^{Diff}(Q^2,Y,y) \approx {\bf \frac 1{N_c^2}}\ 
\frac {{\cal N}}{x_P}  
\ \frac {e^{2y\D}}{4\pi \D ^{''} y}
\ \sqrt {\frac 2
{1\!+\!2\eta}}\n
\!\!\!\!\!\!\!\!\!\!\!\!\! \times
\exp \left\{ (Y\!-\!y)\ {\bf \epsilon}   _s \right\}\ \left(\frac 
Q{Q_0}\right)^{2{\bf \g_s}}
\exp{\left( - \frac {2 \eta}{1\!+\!2\eta}\ \frac {2\log^2\left(\frac 
Q{Q_0}\right)}{\D ^{''} 
(Y\!-\!y)}\right)} \ ,
\l{biczy10}
\ea
where $\g_s$ and $\epsilon_s$ are defined  in (\ref{epsilon}) and a {\it known} 
normalization
\be
{\cal N} \equiv \vert\D^{'}(\frac {1\!-\!\g_c}2)\!+\!\D^{'}(\g_c)\vert\  
h(\g_c) \ I_{T,L}(\g_c) \times {\cal N}^{tot}\ ,
\l{norm}
\ee
with $\g_c$ as in (\ref{biczy9}).
 Note that we could also have considered $\g_s (y)$ instead of the constant 
$\g_c$ in formula (\ref{norm}) since 
the 
prefactors are expected to be 
slowly varying functions of $\g.$ In fact, care must be taken about  poles at 
$\g=0$ 
contained in the 
functions $I_{T,L}(\g)$ (the same poles appear consistently in both sides of 
(\ref{ident}), as 
explicit in the definition (\ref{Pf}) of $P\!f_{T,L} (\g)$). A normalization 
correction could appear in
(\ref{biczy9},\ref{biczy10}) by a 
known rescaling factor \cite{chris}. We leave a  detailed phenomenological study 
for futher work.

The relations (\ref{biczy9},\ref{biczy10}) call for some comments. On a 
theoretical point of 
view, these 
relations are out of 
range of the 
present knowledge of QCD, for its large distance, confining regime. However, 
some arguments are in favor of such kind of relations. For instance, there 
exists  
a 
complementarity between  the perturbative results and   the 
non 
perturbative ansatze present here. The value of the strong coupling 
constant $\a,$ 
the 
number of primordial dipoles $n_{eff}$ and, in the derivation \cite {bi98} of 
hard 
diffraction, the expression of the cut-off $\Theta$  are not determined in the 
BFKL approach 
which essentially means that  the relative normalization of diffractive vs. 
non-diffractive 
contributions remain unknown. It is just this ratio
 which is determined by the Soft Colour Interaction ans\"atz. 
On 
the other hand,
the non-trivial form of the structure functions inherited from the QCD dipole 
calculations 
could not be guessed just from what we know about soft colour processes, which 
are 
limited by  our present lack of knowledge of the theory of strong interactions 
at long 
distances. 
 \section{Summary and outlook}
 Summarizing  our results, let us quote:
 
 {\bf 1)} Using S-Matrix properties of triple-Regge contributions, a relation is 
found 
between  three popular approaches to hard 
diffraction: the 
partonic Pomeron, Soft Colour Interaction and QCD dipole formulations.
 
 {\bf 2)} A  formulation of the Soft Colour Interaction approach is 
proposed as a 
specific 
double discontinuity of a $3 \to 3$ forward amplitude.

 {\bf 3)} The ``effective'' parameters of 
the $\g_c$-Pomeron 
total cross-section corresponding to the partonic Pomeron are determined from  
leading 
log perturbative  QCD 
resummation. They are found 
to depend not only on on  $Q^2$ but also on rapidities through the parameter 
$\eta = 
\log (1/\beta)/\log (1/x_P)\approx <\log 
(M_X^2)>/y_{gap}.$
 
 {\bf 4)} The diffractive structure functions $F_{T,L}^{Diff}(Q^2,Y,y)$ are 
found to be 
determined both in 
form and normalization, using their perturbative (through QCD dipoles) and non 
perturbative 
(through 
Soft Color Interaction) relationship with the total structure functions 
$F_{T,L}^{tot}(Q^2,Y,y).$
 
 Let us add a few comments concerning these points. 
 
 Concerning the {\it partonic Pomeron} approach, we find\footnote{This remark 
was made 
in Ref.\cite{ba95} about the BFKL result and is extended here to the partonic 
Pomeron.} 
that  the ``effective'' 
parameters of  
$\sigma^{(T,L)/tot}_{\gamma^*-P}$ depend on a parameter $\eta$ external to the 
$\gamma^*-P$ 
reaction. Looking to the deep reason of this fact, we find the constraint 
$\g_1+\g_2+\g =1$  which 
prevents the BFKL Pomeron singularities of the involved BFKL Pomerons from 
taking their ordinary 
saddle-point positions. Hence those Pomerons responsible for the gaps become 
$Q^2$-independent 
while 
the one in $\sigma^{(T,L)/tot}_{\gamma^*-P}$ becomes both $Q^2$ and 
$\eta$-dependent.  This kind 
of 
``off-shellness'' is quite remarkable, showing that a universal Pomeron kernel 
$\D(\g)$ can lead 
to 
non-universal effective  behaviours when put in a hard interaction context. It 
would be 
interesting  to look for an experimental study of this intringuing feature of 
the 
partonic Pomeron. The compatibility with DGLAP evolution equations is also to be 
studied 
in detail.
 
 Concerning the {\it Soft Colour Interaction} prediction, relating the 
diffractive and 
non-diffractive 
components of deep-inelastic scattering, it seems quite compatible with the QCD 
predictions on the 
form of the amplitudes. However, an unexpected difference appears, since the 
typical values of the 
``effective'' dimensions are different in both cases. it is clear for the 
integrated diffractive 
contribution  $F_{T,L}^{Diff/tot}$, see (\ref{biczy4}), comparing $\g_{s.p.}$ 
and  $\g_c.$ It is 
also 
explicit from the analysis of $F_{T,L}^{Diff}$ itself that the relevant regions  
in the inverse 
Mellin transforms are always for values of $\g$ lower than for $F_{T,L}^{tot}.$ 
It will be 
interesting to test this prediction with data.
 
For developping this study beyond the first theoretical step, we would like to 
stress the feasability of   phenomenological tests and applications. However 
this requires 
a 
similar analysis for 
the 
``elastic'' QCD dipole component. One knows that the Mueller Regge formulae can 
be extended to 
generalized Pomeron Regge amplitudes (e.g. \cite{sa74}). We find interesting to 
elaborate on the 
S-matrix discontinuity properties in this context and their possible relations 
with 
``higher-twist'' 
QCD contributions which are known to be relevant in this case. This deserves a 
specific study 
which 
is beyond the scope of this paper. 

On the theoretical ground, some interesting problems arise. For instance, the 
known problem of the application of a (resummed) perturbative scheme at $t \ne 
0$ require a better understanding of the matching between perturbative and 
non-perturbative features. Also are interesting to develop the more 
precise extension of 
this scheme to Regge cuts,  next-leading BFKL contributions and, last but not 
least, 
unitarity corrections, which are required to identify properly the soft Pomeron 
contributions. Finally, the 
study of Tevatron 
results on diffraction will be welcome, since the Soft Colour Interaction 
approach seems to give some interesting 
clues, 
while factorization of the parton structure functions of the Pomeron appears to 
be violated.

 \vspace{0.3cm}
{\bf Acknowledgements}
\vspace{0.3cm}

Thanks are due to Jochen Bartels, who draw the attention of one of us 
(R.P.) to the 
relations betwen Regge discontinuities cf. the papers \cite {ba95,ba96},  to 
Andrzej 
Bialas for fruitful discussions and  
the inspiring paper in collaboration with Wieczlaw  Czyz
\cite{bi98} , and to Christophe Royon, for 
his help 
and advice.
\eject
{\bf APPENDIX: Saddle-point calculation of section 4}
\\
 \\
 Starting with Eq. (\ref{biczy4}), the integration over $\g_1-\g_2$ leads to
\be
F_{T,L}^{Diff}= 
{\cal N}^{Diff}\frac {<P_T^2>}{x_P Q_0^2}\ \int_{c-i\infty}^{c+i\infty}\frac 
{d\g}{2i\pi}\sqrt {\frac 
1 {4\pi \D ^{''}\! (\frac {1-\g} 2)\ 
y}} 
\Theta P\!f_{T,L} \ 
\exp  {\cal H}(y,\g)\ ,
\l{biczy11}
\ee
where the function
\be
{\cal H}(y,\g)\equiv\left\{2y \D\left(\frac {1\!-\!\g} 2\right) + (Y\!-\!y)\ 
\D(\g) + 2\g 
\log\frac 
Q{Q_0}\right\}
\l{debut}
\ee
determines the saddle-point equations, namely
\ba
\frac{\partial{\cal H}(y,\g)}{\partial\g} &\equiv& -y\ \D^{'}\left(\frac 
{1\!-\!\g} 2\right) + (Y\!-\!y)\ \D^{'}(\g) + 2
\log\frac 
Q{Q_0}=0
\n
\frac{\partial{\cal H}(y,\g)}{\partial 
y}&\equiv&2\D\left(\frac{1-\g}{2}\right)-\D(\g)=0\ ,
\l{cols}
\ea
whose solution $(y_c,\g_c)$ is given in (\ref{solution}), see the text.

The prefactors appearing in the integrated formula are coming from the partial 
second derivatives of ${\cal H}(y,\g).$ A first one comes from the partial 
second derivative over $\g$
\be
\frac 1{\sqrt{2\pi  \ \frac{\partial^2{\cal H}}{\partial\g^2}(y_c,\g_c)}}=\frac 
1{\sqrt{2\pi \left[\frac {y_c}2 \D^{''}\left(\frac {1\!-\!\g_c} 2\right) + 
(Y\!-\!y_c)\ \D^{''}(\g_c)\right]}}\ .
\l{pref1}
\ee
The second one is slightly more involved. It reads
\be
{\sqrt\frac {2\pi }
{\frac{\partial{\g_c}}{\partial y}
\left[2 \frac{\partial^2{\cal H}}{\partial\g \partial y} +\frac{\partial^2{\cal 
H}}{\partial\g^2}\frac{\partial{\g_c}}{\partial y}
\right]
}}
=\sqrt{\frac {2\pi \left[\frac {y_c}2 \D^{''}\left(\frac {1\!-\!\g_c} 2\right) + 
(Y\!-\!y_c)\ \D^{''}(\g_c)\right]}
{\left(\D^{'}(\frac {1\!-\!\g_c}2)\!+\!\D^{'}(\g_c)\right)^2}}\ .
\l{pref2}
\ee
The simplification appearing in the second prefactor comes from an identity
\be
\frac{\partial^2{\cal H}}{\partial y^2}\frac{\partial{\g_c}}{\partial 
y}+\frac{\partial^2{\cal H}}{\partial\g \partial y}=0
\l{identity}
\ee
at the saddle-point $(y_c,\g_c),$ which came from the calculation.

Noting that the numerator of (\ref{pref2}) simplifies with the denominator of 
(\ref{pref1}),  we are led to the final remarkably simple formula (\ref{biczy7}) 
of the text.

\end{document}